\documentclass{PoS}
\usepackage{amsmath,slashed,subfigure}
\pdfoutput=1
\newcommand{\su}[1]{\ensuremath{\text{SU}(#1)}}
\newcommand{\so}[1]{\ensuremath{\text{SO}(#1)}}
\newcommand{\uu}[1]{\ensuremath{\text{U}(#1)}}
\newcommand{\bpsi}{\bar{\psi}}
\DeclareMathOperator{\Tr}{Tr}

\title{Lattice simulations of technicolour theories\\ with adjoint fermions
and supersymmetric Yang-Mills theory}

\ShortTitle{Lattice simulations of adjoint QCD}

\author{\speaker{Georg Bergner}\\
        Universit\"at Bern, Albert Einstein Center for Fundamental Physics,\\
        Institut f\"ur Theoretische Physik,  Sidlerstr.~5,
        CH-3012 Bern, Switzerland\\
        E-mail: \email{bergner@itp.unibe.ch}}

\author{Pietro Giudice, Gernot M\"unster\\
Universit\"at M\"unster, Institut f\"ur Theoretische Physik,\\
Wilhelm-Klemm-Str. 9, D-48149 M\"unster, Germany\\
E-mail: \email{p.giudice@uni-muenster.de, munsteg@uni-muenster.de}}

\author{Istvan Montvay\\
Deutsches Elektronen-Synchrotron DESY,\\
Notkestr. 85, D-22603 Hamburg, Germany\\
E-mail: \email{montvay@mail.desy.de}}

\author{Stefano Piemonte\\
Universit\"at Regensburg, Institute for Theoretical Physics,\\
D-93040 Regensburg, Germany\\
E-mail: \email{stefano.piemonte@ur.de}}

\abstract{Theories with fermions in the adjoint representation have several interesting applications in extensions of the standard model. 
The conformal window for these theories is of particular interest for technicolour extensions. We present here our newest results for the spectrum of $N_f=2$ adjoint QCD and
compare them with the predictions for a conformal behaviour. The comparison with supersymmetric Yang-Mills theory, investigated with the same methods, will help to distinguish more
clearly the conformal and the confining scenario. The spectrum includes additional fermionic states that are not present in QCD. We provide results for the mass of these states and discuss 
their phenomenological relevance. In addition we have done preliminary investigations of the singlet scalar meson state.}

\FullConference{The 33rd International Symposium on Lattice Field Theory\\
		14 -18 July 2015\\
		Kobe International Conference Center, Kobe, Japan*}

\begin{document}

\section{Introduction}
The investigation of SU(N) gauge theories with fermions in higher representation has several different motivations.
Some of them are of pure theoretical nature like the question whether a dynamics and a particle spectrum completely different 
from QCD can be observed.

Even more interesting are phenomenological applications of these theories. In possible extensions of the standard model they
are an alternative for simply modified versions of QCD with fermions in the fundamental representation. One example are 
technicolour theories that provide a more natural representation of the electroweak sector by introducing a new strong dynamics.
The Higgs particle emerges in this case as a bound state in the new strongly interacting sector.

There are different requirements from phenomenological models for a technicolour theory that depend on the chosen extension of the standard model.
In this work we focus on the near conformal behaviour, the  appearance of a light scalar particle, 
and a large mass anomalous dimension. 
The near conformal, or walking, behaviour with a small number of fermion flavours is required to
avoid possible tensions with electroweak precision data. This can be achieved with fermions in higher representation of the gauge group. 
Several analytical \cite{Sannino:2004qp,Dietrich:2006cm,Braun:2010qs} and numerical lattice studies
\cite{Catterall:2007yx,Hietanen:2009zz,DelDebbio:2010hx,DeGrand:2011qd,Appelquist:2011dp,DeGrand:2010na,Fodor:2015zna,Hasenfratz:2015ssa,Athenodorou:2014eua}
have been dedicated to the investigation of the conformality in different gauge theories.

The conformal behaviour manifests itself in the mass spectrum of the theory. All states $M$ should scale to zero according to $M\propto m^{1/(1+\gamma_\ast)}$ with the residual quark mass $m$
and the same mass anomalous dimension $\gamma_\ast$. This behaviour should be observable if $m$ is below a certain threshold. It is much different from the chiral symmetry breaking scenario, 
where a clear separation between the pseudo Nambu-Goldstone bosons (pNGb) and the rest of the spectrum appears at small $m$ and eventually the mass of the pNGb goes to zero in the chiral limit, 
whereas the mass remains finite for the other particles. It is in general difficult to discern to which of the two classes a considered theory belongs since one is always restricted to a certain range of $m$ in the lattice simulations and the chiral limit can only be extrapolated. A comparison of different theories might therefore help to resolve the differences between conformal and chiral symmetry breaking scenario.
It is important to choose a comparable lattice realisation in such a comparison since lattice artefacts might have a significant influence on the scaling behaviour. 

The adjoint representation is particularly interesting among the higher representations of the gauge group.
The minimal walking technicolour (MWT), the \su{2} gauge theory with two Dirac fermions in the adjoint representation, is a candidate for a technicolour extension of the 
standard model. Further interesting gauge theories with fermions in the symmetric and anti-symmetric representation are related to the adjoint representation by large $N_c$ equivalence.
This leads to constraints for the conformal window of the symmetric representation that can be deduced from the adjoint one \cite{Bergner:2015dya}.
In gauge theories with fermions in the adjoint representation, the number of degrees of freedom in the fermionic and bosonic sector can be equal, as required by supersymmetry.
Therefore supersymmetric Yang-Mills theory (SYM) is also among the strongly interacting gauge theories having fermions in the adjoint representation (adjoint QCD).
Specific states that appear in the adjoint representation, but not in the fundamental one, might have applications in extensions of the standard model.

%In this work we present a comparison of \su{2} gauge theories with a different number of fermions in the adjoint representation. It includes some preliminary data from new simulations, in particular of
%MWT. We will compare SYM, a confining theory, and MWT for which a conformal behaviour has been found in also in previous lattice simulations.
%For the MWT we will show for the first time results for a specific spin-$1/2$ state that can be constructed in the adjoint representation. We will also discuss the measurement of the 
%scalar state using mesonic correlators with disconnected contributions.

\section{Continuum action and chiral symmetry breaking in adjoint QCD}
The Lagrangian of adjoint QCD has the following form
\begin{equation}
 \mathcal{L}=
 \Tr\left[-\frac{1}{4}F_{\mu\nu}F^{\mu\nu}
 +\sum_{i}^{N_f}\bpsi_{i}(\slashed{D}+m)\psi_{i}\right]\, ,
\end{equation}
where we assume the gauge symmetry to be \su{2}. 
$\psi$ is a  Dirac-Fermion in the adjoint representation
with the covariant derivative 
\begin{equation}
D_\mu \psi =\partial_{\mu}\psi + i g [A_{\mu},\psi]\, .
\end{equation}

The adjoint representation is consistent with the Majorana condition
$\lambda=C\lambda^T$ and each Dirac fermion can be decomposed into two Majorana fermions.
$N_f$ counts the Dirac flavours and, consequently, there are theories with half integer $N_f$
corresponding to an odd number of Majorana flavours.
The representation in terms of $2N_f$ Majorana flavours indicates a chiral symmetry breaking pattern by the formation of a chiral condensate that is different from QCD:
\begin{align}
 \su{2N_f} \rightarrow \so{2N_f}\, .
\label{chiralbreak}
\end{align}
Consequently there are  $2N_f^2+N_f-1$ pNGb in adjoint QCD.

Notably there are a number of different notations for the pNGb and other states of these theories.
This comes from the different context in which these theories are considered, related to supersymmetry, QCD, or the \su{2}
version of QCD. The symmetry breaking pattern of \eqref{chiralbreak} cannot be directly realised in SYM with $N_f=1/2$, but the theory can be considered as partially quenched 
$N_f=1$ adjoint QCD. The corresponding partially quenched chiral perturbation theory has been formulated in \cite{Munster:2014cja} and is applied
in the extrapolation of the chiral limit in SYM. In this case the pNGb is called adjoint pion to emphasise the similarity to
chiral perturbation theory in QCD. 

In $N_f=1$ adjoint QCD the unbroken \so{2} is equivalent to the $\uu{1}_V$ corresponding to the Baryon number conservation in the Dirac fermion formulation. The 
\su{2} gauge group allows to construct baryonic operators from two fermion fields. In \cite{Athenodorou:2014eua} the pNGb, represented by the $\psi^T C \gamma_5 \psi$,
has therefore baryon number $2$ and is called scalar baryon.
In the investigations of MWT the pNGb is usually called pseudoscalar meson \cite{DelDebbio:2010hx}.
\section{Lattice setup}
In our simulations we have chosen a tree level Symanzik improved gauge action and a Dirac-Wilson operator with stout smeared links in the fermionic part of the action.
More details of our results for SYM have been presented in other contributions of this conference \cite{Bergner:2015cqa,Bergner:2015lba}. For this theory we have data at 
three different $\beta$ in the range of $1.6$ to $1.9$, which allows a reasonable estimation of lattice artefacts and continuum extrapolations.
The mass of the adjoint pion in lattice units in the relevant runs varies from $0.6$ down to $0.2$.

Apart from the different number of fermion flavors, the same lattice action was used in the simulations of MWT. The considered range of $\beta$ values is in this case limited by the bulk transition.
We have determined a bulk transition around $\beta=1.4$ with our lattice action. The control of finite volume effects is important  in the 
investigations of a conformal theory. Therefore we have chosen a rather small $\beta$ of $1.5$ in our first analysis. In a second step we have also 
done simulations at $\beta=1.7$ to check for possible lattice artefacts. The pion mass at these runs was between about $0.9$ and $0.2$ in lattice units.
%%%%%%%%%%%%%%%%%%%%%%%%%%%%%%%%%%%%%%%%%%%%%%%%%%%%%%%%%%%%%%%%%%%%%%%%%%%%%%%%%%
\section{Conformal window and comparison of MWT and SYM}
\begin{figure}[h]
\centering
\subfigure[SYM $0^{++}$ extrapolation]{\includegraphics[width=0.48\textwidth]{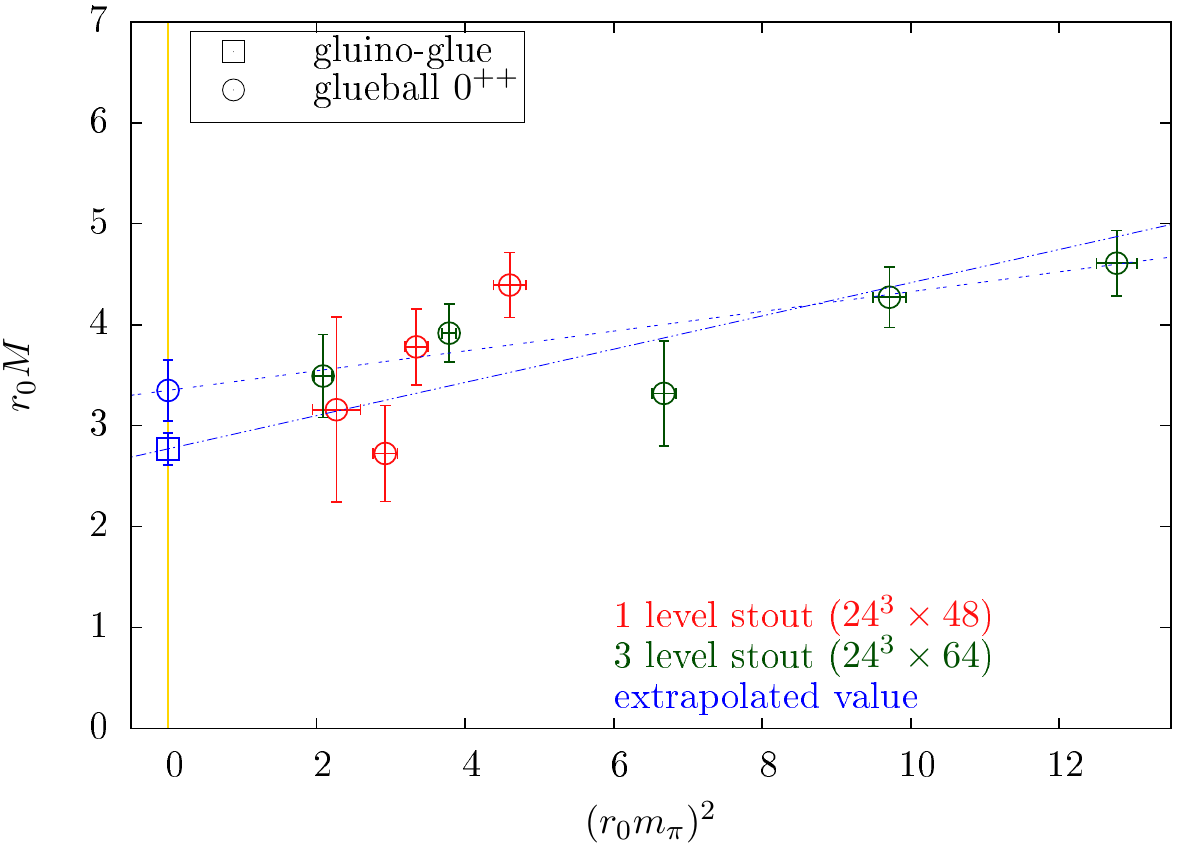}\label{gbextrapol}}
%\hspace*{-0.1cm}% solves bug in my pdftex version
\subfigure[SYM $\text{a-}f_0$ extrapolation]{ \includegraphics[width=0.48\textwidth]{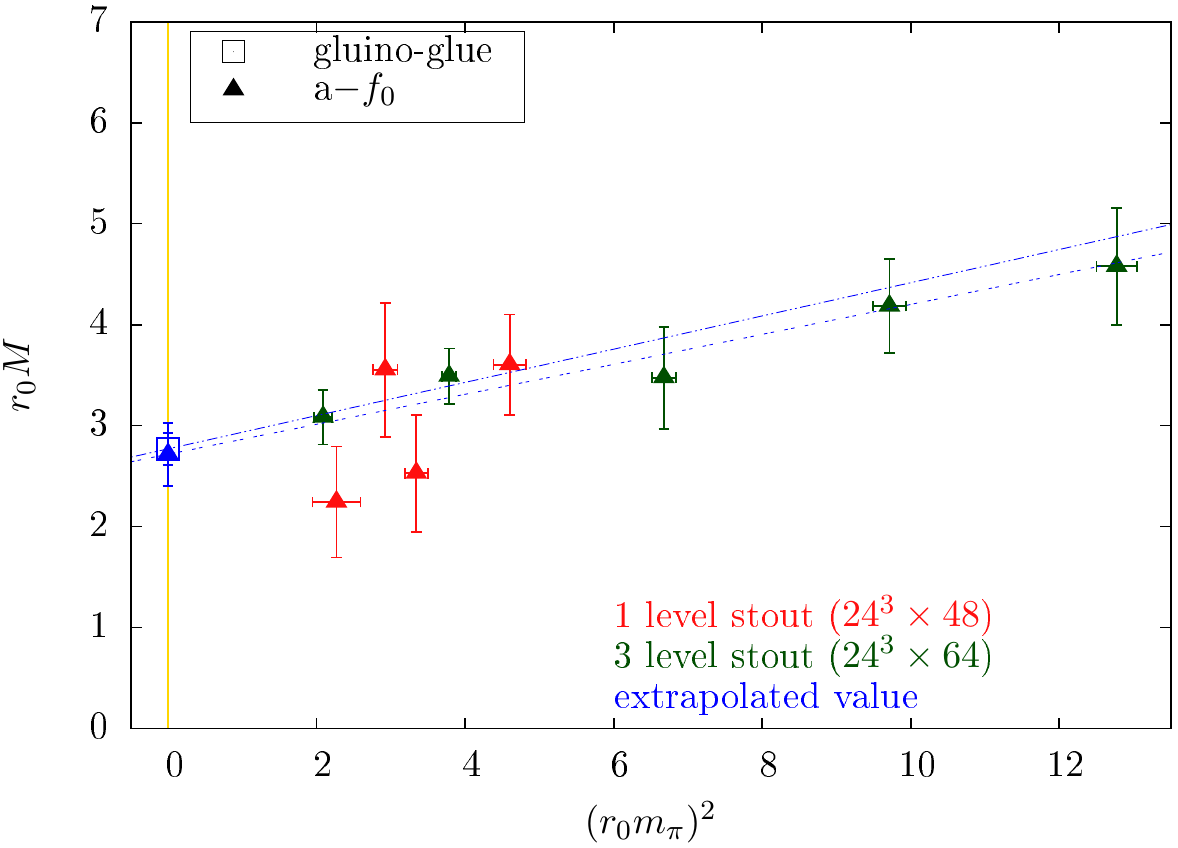}\label{f0extrapol}}
 \caption{This figure shows a part of the mass spectrum of supersymmetric Yang-Mills theory at $\beta=1.75$ and the extrapolation to the chiral limit defined by a vanishing adjoint pion mass $m_{\pi}$. 
  For the gluino-glue particle only the result of the chiral extrapolation, but not the measured data points are shown.
 (a) The fermionic gluino-glue and the bosonic scalar glueball $0^{++}$ are compared.  (b) The scalar singlet meson, $\text{a-}f_0$, is shown in comparison with the chiral extrapolation of the gluino-glue particle. }
\end{figure}

Our results for the mass spectrum of SYM have been reported in \cite{Bergner:2013nwa,Bergner:2015lba}. We have performed extrapolations to the chiral limit defined by a vanishing adjoint pion mass at each fixed lattice spacing.
At this chiral point, supersymmetry, which is broken by the lattice regularisation, is restored in the continuum limit, and even at a finite lattice spacing
we find no considerable indication for supersymmetry breaking in the supersymmetric Ward identities.
The extrapolation in the scalar sector is shown in Fig.~\ref{gbextrapol} and \ref{f0extrapol}. The masses of the other states are larger than the adjoint pion mass in 
the considered parameter range and they extrapolate to a finite value in the chiral limit. As required by supersymmetry there is a degeneracy between bosonic and fermionic masses.
The scalar singlet meson, $\text{a-}f_0$, and the $0^{++}$ glueball have almost the same mass. These operators have the same quantum numbers and seem to have both a reasonable overlap with the 
ground state in this channel. 

\begin{figure}[t]
\centering
\subfigure[MWT masses in lattice units]{\includegraphics[width=0.48\textwidth]{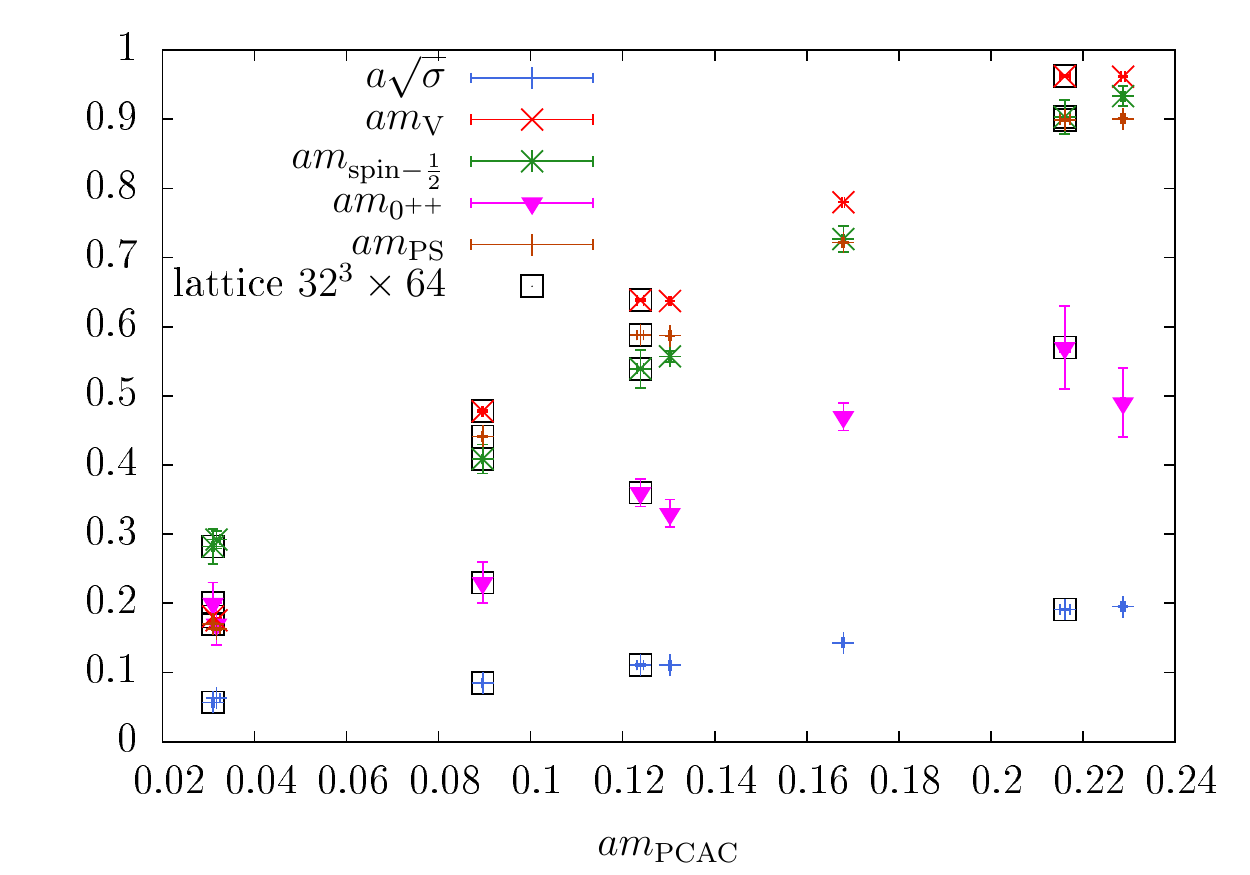}\label{mwtspect1}}
%\hspace*{-0.1cm}% solves bug in my pdftex version
\subfigure[MWT mass ratios]{ \includegraphics[width=0.48\textwidth]{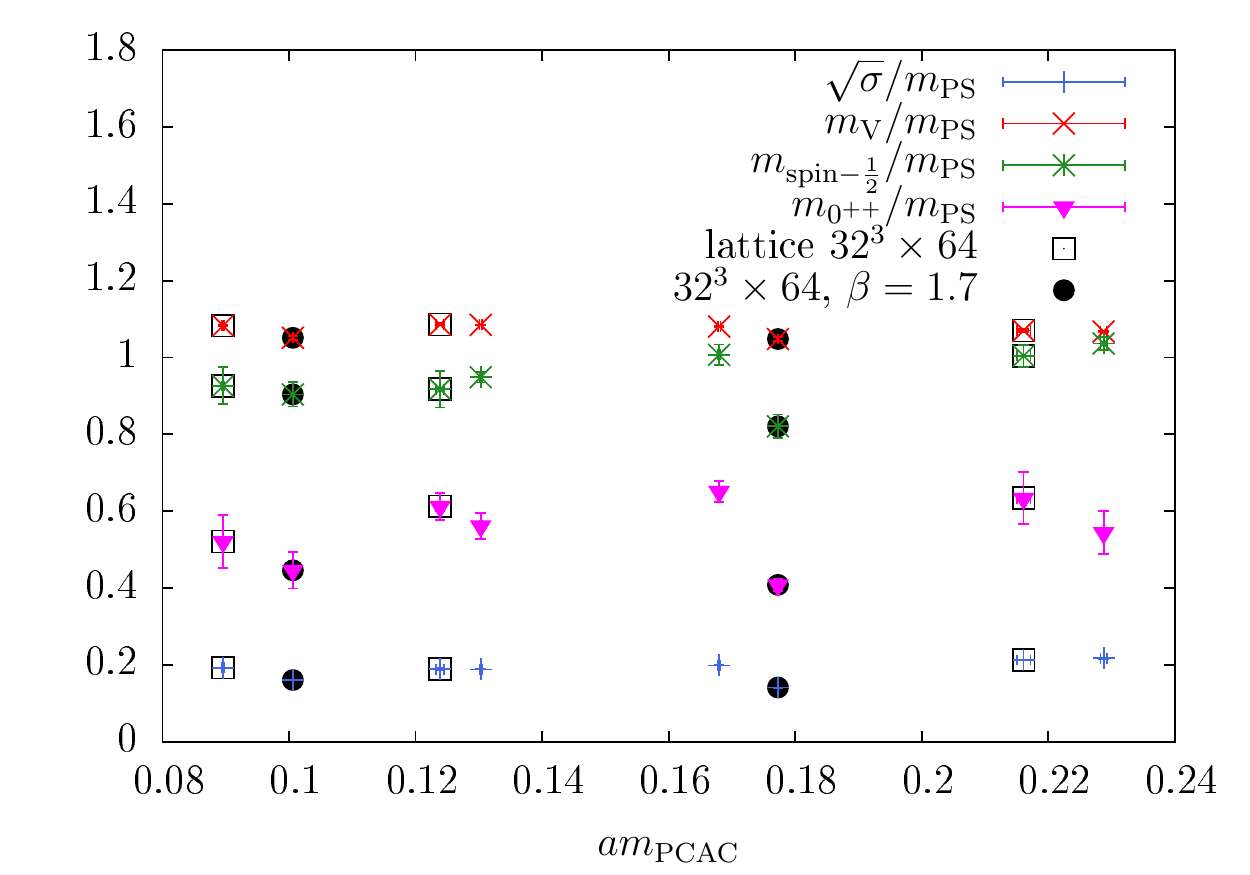}\label{mwtspect2}}
 \caption{The particle masses in $N_f=2$ adjoint QCD (MWT) are shown as a function of the PCAC quark mass ($m_{\text{PCAC}}$). (a) The masses of the vector ($m_\text{V}$) and pseudoscalar ($m_{\text{PS}}$) meson, the spin-1/2 state ($m_{\text{spin-1/2}}$), and the $0^{++}$ glueball are shown together with the string tension $\sigma$ for the two different volumes $24^3\times 64$ and $32^3\times 64$ ($\beta=1.5$). All quantities are represented in lattice units. (b) The mass ratios of these different states and the pseudoscalar meson mass excluding the runs with the smallest $m_\text{PCAC}$ that are probably affected by finite size effects. The data at the smaller lattice spacing ($\beta=1.7$) are included in this figure. }
\end{figure}

The results for MWT are completely different, see Fig.~\ref{mwtspect1} and \ref{mwtspect2}. The chiral extrapolation is in this case determined by the PCAC quark mass.
All of the masses scale to zero in the chiral limit and their ratios are constant. The particle with the lowest mass is the scalar glueball and not the pseudscalar meson, the pNGb in this theory.
Our simulations at the second $\beta$, corresponding to a smaller lattice spacing, are consistent with this picture.
These observations are clearly indicating a conformal scenario, in contrast to the chiral symmetry breaking scenario of SYM.
As shown in Fig.~\ref{mscaling}, we find a scaling of the states with a mass anomalous dimension around $0.38$ which is consistent with what has been determined in \cite{Debbio:2014wpa} for this theory. 
There is a deviation at the two lightest PCAC masses where the mass of the spin-1/2 and the glueball become heavier than the mesons, but even at the $32^3\times 64$ lattice it is most likely a finite size effect. 
In general the finite size effects are severe in MWT as was also pointed out in \cite{Debbio:2014wpa} and we plan to clarify the relevance of these effects in our next investigations.

\section{Fractionally charged particles and scalar singlet meson operators in MWT}
The adjoint representation of the fermions allows for certain states that have no counterpart in QCD with fundamental fermions. One of them is a spin-1/2 operator 
composed of fermion fields and gauge bosons
\begin{equation}
  O_{\text{spin-1/2}}=\sum_{\mu,\nu} \sigma_{\mu\nu} \Tr\left[F^{\mu\nu} \lambda \right]\, .
\end{equation}
In SYM this operator is essential since it corresponds to the gluino-glue, the fermionic partner of the bosonic glueball or meson operator. Unbroken supersymmetry implies multiplets of 
fermions and bosons with the same mass. The low energy effective theory must therefore contain such kind of fermionic bound states.

In MWT the spin-1/2 state is relevant for phenomenological considerations since it leads to fractionally charge particles when a naive hypercharge assignment is used. Even though the 
mass of these particles has not been measured before, they have been considered to disfavour the phenomenological relevance of the theory. This was essentially one of the motivations
to consider \so{4} gauge theory as an alternative \cite{Hietanen:2013gva}. In \cite{Kouvaris:2007iq} the particles were, on the other hand, considered as an alternative dark matter scenario.
With our current investigations we were able to show that the mass of the spin-1/2 is well separated from the lightest scalar particle in the theory. On the other hand, it is slightly lighter than the 
pseudoscalar meson, which means that it could be one of the first experimentally observable ``new physics'' states in this theory.

\begin{figure}[t]
\subfigure[MWT scaling of the masses]{\includegraphics[width=0.48\textwidth]{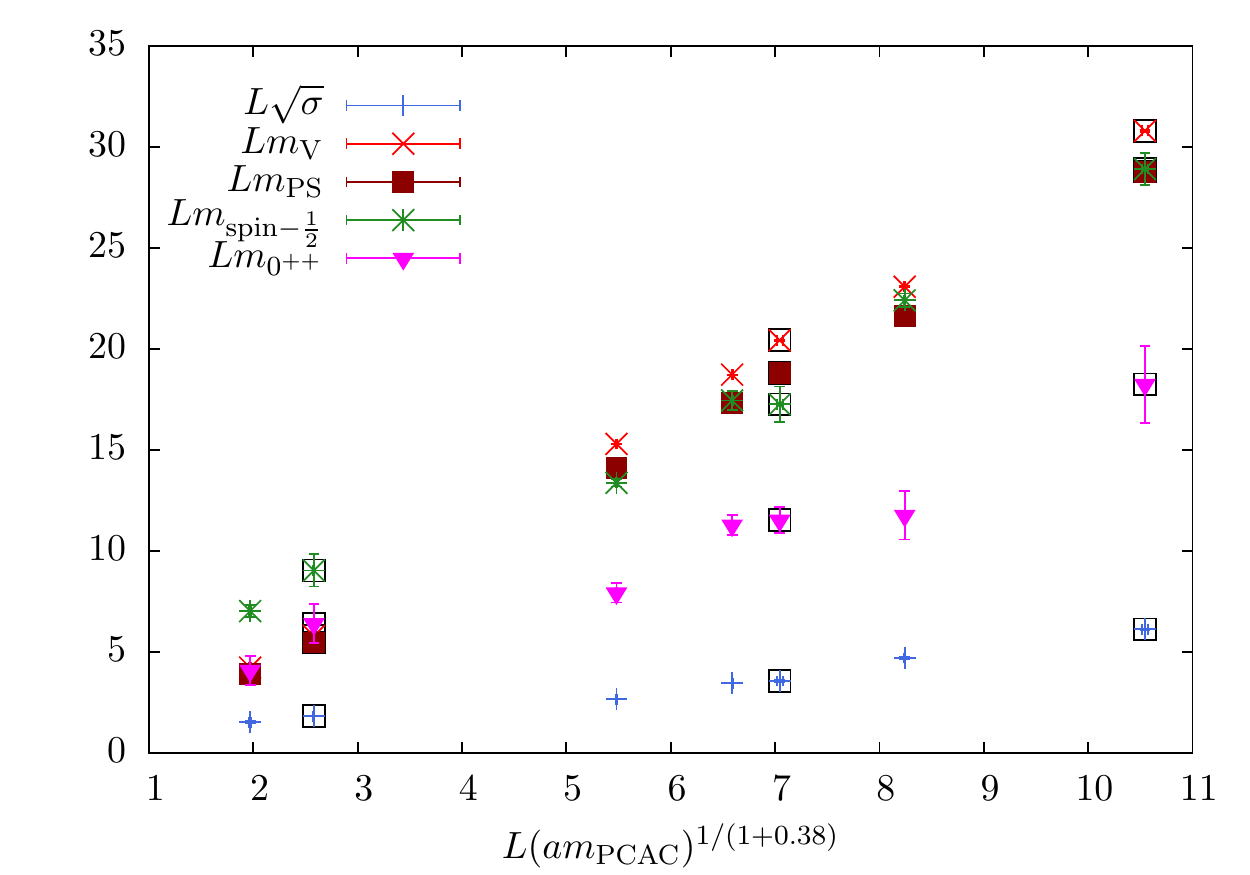}\label{mscaling}}
%\hspace*{-0.1cm}% solves bug in my pdftex version
\subfigure[MWT scalar meson contributions]{ \includegraphics[width=0.48\textwidth]{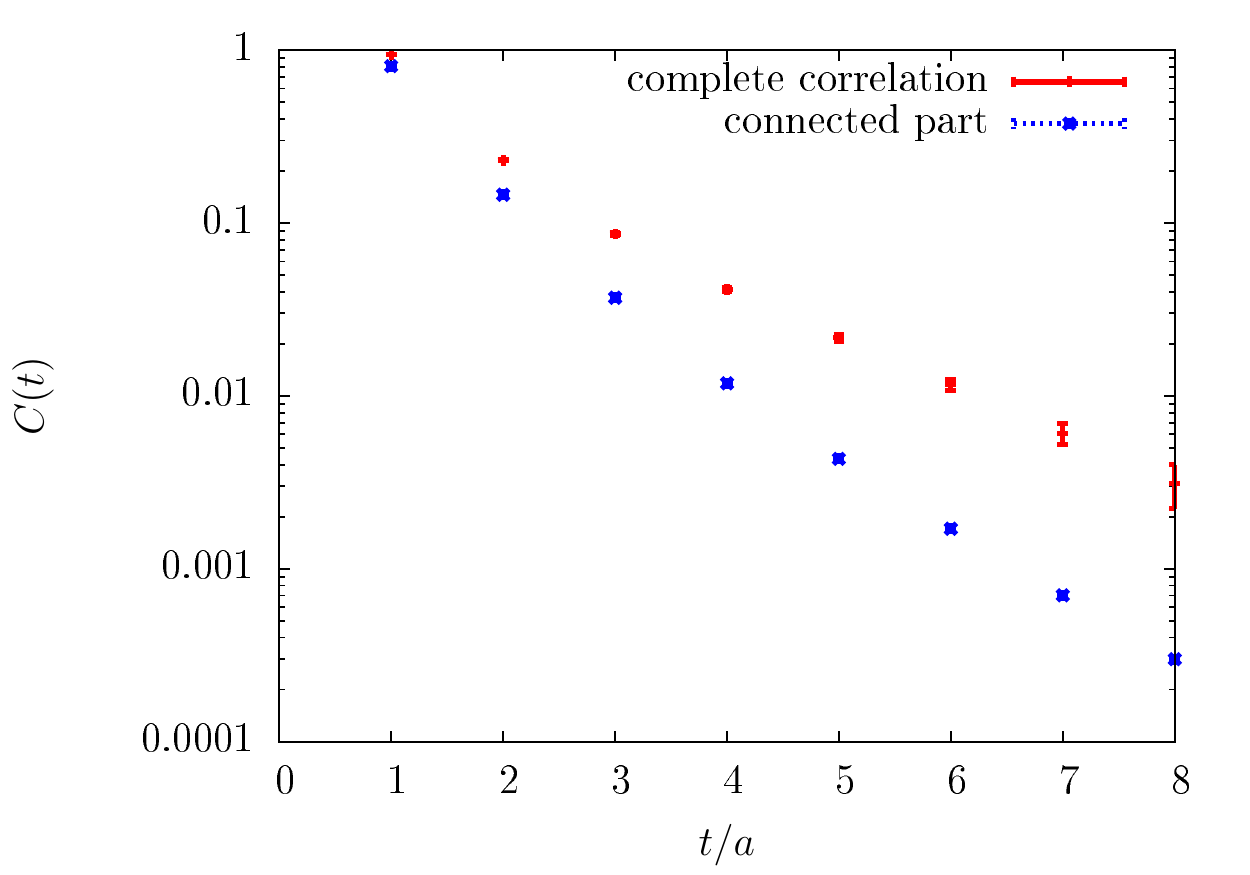}\label{f0disc}}
\caption{(a) The scaling of the masses compared to the expected scaling with a mass anomalous dimension of $0.38$. A linear behaviour in this plot indicates that the data are consistent with the conformal scaling.
(b) An example for the connected and disconnected contributions to the scalar singlet meson correlator $C(t)$ at $\beta=1.5$, $m_\text{PS}=0.5873(3)$. }
\end{figure}

The light scalar state is one of the most important ingredients of this theory since it explains the observed Higgs particle. In SYM the 
scalar singlet meson operator and the glueball provide two independent measurements of the lightest scalar mass (compare Fig.~\ref{gbextrapol} and \ref{f0extrapol}).
For this reason we have measured besides the glueball also the scalar singlet meson in MWT. 
The correlator of the scalar meson is dominated by the disconnected contribution, as shown in Fig.~\ref{f0disc}, resulting in a large separation between the scalar singlet and the triplet meson.
In this particular example  the mass in lattice units of the triplet (disconnected) is $0.747(17)$ and the one of the singlet (connected + disconnected) is $0.540(53)$. The obtained scalar mass is below 
the pseudoscalar meson mass ($0.5873(3)$). We were, however, not able to see the degeneracy with the light scalar glueball which has a mass of only $0.33(2)$ in lattice units. One reason is, of course, the large separation of this state from the rest of the spectrum, but it might also be a more generic feature of conformal theories.

\section{Comparison to one flavour adjoint QCD and outlook}
MWT and SYM show clear indications for a much different behaviour. SYM is expected to be a confining theory, which is in accordance with the lattice results. For MWT we find the indications of a conformal behaviour, in accordance with other lattice studies.
Between these two examples with a reasonable signal for a non-conformal and conformal behaviour there are the theories with $N_f=1$ and $N_f=3/2$. The data from the first lattice simulations of $N_f=1$ have been published in \cite{Athenodorou:2014eua}. There is evidence that even this theory is rather on the conformal side in terms of, for example, the nearly constant mass ratios. Further features of the theory are a light scalar and a large mass anomalous dimension of $\gamma_\ast=0.92(1)$. In contrast to MWT the spin-1/2 particle is clearly heavier than the pNGb and 
there is a degenerate signal for the scalar mass in the singlet meson and the scalar glueball measurement.
In further studies we will complete the picture with results for $N_f=3/2$ adjoint QCD. 
The determination of the conformal window for the adjoint representation has interesting consequences also for the studies of other theories, in particular with fermions in the symmetric representation. 
MWT appears to be one of the most challenging theories from the point of view of numerical simulations due to the large finite volume effects.
%%%%%%%%%%%%%%%%%%%%%%%%%%%%%%%%%%%%%%%%%%%%%%%%%%%%%%%%%%%%%%%
\section*{Acknowledgments}
This project is supported  by the John von Neumann Institute for Computing
(NIC) with grants of computing time. 
We also gratefully acknowledge the Gauss Centre for Supercomputing e.V. for funding this project by providing 
computing time on the GCS Supercomputer SuperMUC at Leibniz Supercomputing Centre.
Further computing time has been
provided by the compute cluster PALMA of the University of M\"unster.
%\begin{spacing}{0.9}

%\end{spacing}
\end{document}